\documentclass[prl,twocolumn,showpacs,amsmath,amssymb]{revtex4}
\usepackage{graphicx}
\usepackage{dcolumn}
\usepackage{bm}

\begin{document}


\title{Conductance behavior with temperature and magnetic field in the disordered films of titanium nitride}
\author{P. Kulkarni$^{*}$}
\affiliation{Laboratorio de Bajas Temperaturas, Departamento de F\'isica de la Materia Condensada, Instituto de Ciencia de Materiales Nicol\'as Cabrera, Facultad de Ciencias, Universidad Aut\'onoma de Madrid, 28049 Madrid, Spain}
\author{H. Suderow}
\affiliation{Laboratorio de Bajas Temperaturas, Departamento de F\'isica de la Materia Condensada, Instituto de Ciencia de Materiales Nicol\'as Cabrera, Facultad de Ciencias, Universidad Aut\'onoma de Madrid, 28049 Madrid, Spain}
\author{J.Rodrigo}
\affiliation{Laboratorio de Bajas Temperaturas, Departamento de F\'isica de la Materia Condensada, Instituto de Ciencia de Materiales Nicol\'as Cabrera, Facultad de Ciencias, Universidad Aut\'onoma de Madrid, 28049 Madrid, Spain}
\author{S. Vieira}
\affiliation{Laboratorio de Bajas Temperaturas, Departamento de F\'isica de la Materia Condensada, Instituto de Ciencia de Materiales Nicol\'as Cabrera, Facultad de Ciencias, Universidad Aut\'onoma de Madrid, 28049 Madrid, Spain}
\author{M.R. Baklanov}
\affiliation{IMEC, Kapeldreef 75, B-3001 Leuven, Belgium}
\author {T. Baturina}
\affiliation{Institute of Semiconductor Physics, 13 Lavrentjev Avenue, Novosibirsk, 630090, Russia}
\author{V. Vinokur}
\affiliation{Materials Science Division, Argonne National Laboratory, Argonne, Illinois 60439, USA}

\date{\today}

\begin{abstract}
We report in this paper the temperature and mangetic field dependence of the conductance in the polycrystalline film of titanium nitride, before and after heating at ambient conditions. The difference between the two films is the room temperature sheet resistance which remains within 15 percent and both the films show superconducting transition at lower temperatures. The zero field and the high field data, respectively, corresponds to the superconducting and the normal states. Both the films display Atshuler-Aronov zero bias anamoly in their normal states, and the superconducting gap openeing up at low fields. However the heated film has a smaller gap owing to more pronounced zero bias suppression of the density of states. The normal states in both the films are similar to the quasi-2d-disordered metal and its behavior is studied with temperature. Our data suggests that the zero bias anamoly suppresses the superconducting gap with increase in the disorder. 

\end{abstract}


\maketitle

The interplay between disorder, electron-electron correlations and superconductivity in thin films has been a matter of research for decades, and is still not well understood. A quantum phase transition between a superconductor and an insulator takes place when changing the degree of disorder or the magnetic field\cite{Strongin70,Fisher,Haviland,Hebard}. The resistance per square of the sample is used to quantify disorder, and the transition often takes place around the quantum of resistance at $R_Q\approx h/2e^2$, although the actual value varies in different systems.

Several experiments have addressed the study of local electronic behavior using STM. The superconducting phase of polycrystalline TiN, as well as homogeneously disordered InO and NbN thin films has been analyzed in detail\cite{Noat,Sacepe,Sacepe1,Sacepe2}. The superconducting density of states considerably varies from simple s-wave BCS theory and shows rounded or nearly absent quasiparticle peaks, which have random variations as a function of the position. InO films show a sharply gapped tunneling conductance which loses coherence peaks when closest to the insulator\cite{Sacepe1}. Temperature dependence gives the disappearance of the superconducting quasiparticle peaks at the onset of resistance and the appearance of a pseudogap like background surviving up to temperatures several times T$_c$\cite{Sacepe,Sacepe2}. NbN films, instead, show a filled low energy conductance with very small quasiparticle peaks that change slightly at distances of some hundred nm\cite{Noat}. TiN superconducting polycrystalline ultra thin films show continous distribution of superconducting like tunneling conductance features with quasiparticle peaks whose height is position dependent\cite{Sacepe}. The temperature dependence of the tunneling conductance shows the presence of a precusor pseudogap state\cite{Sacepe,Sacepe1,Sacepe2}. TiN films are particularly interesting, because the superconductor to insulator transition is sharply defined and macroscopic properties are well characterized\cite{Baturina,Baturina1,Pfuner}. Here we make a comparative study of superconducting and normal states in the TiN film using STS measurements at 100 mK.

\begin{figure}[h]
\includegraphics[width=0.48\textwidth]{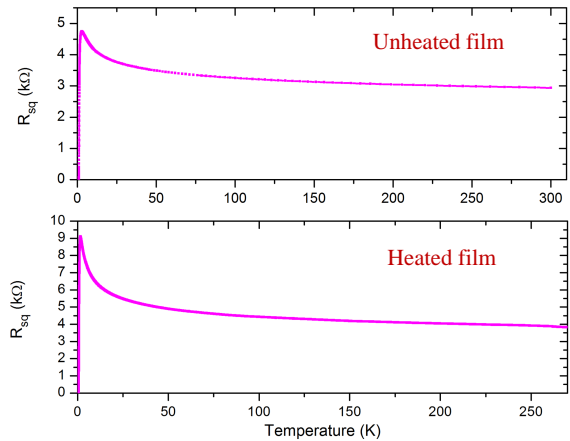}
\caption{The two panels in the figures show the transport behavior in (a) the unheated, and (b) the heated film of TiN. The measurement corresponds to zero magnetic field.}
\end{figure}

We have chosen a polycrystalline TiN thin film prepared using Atomic layer deposition technique. The film was deposited at 350 degree on Si/SiO$_2$ substrate and has thickness of 5 nm. The microstructure of the film shows the crystallites in the size range from 3 to 10 nm, with a mean size of 5 nm. A cross-sectional TEM imaging revealed that the film thickness is the highi of single crystals in the normal direction, and crystallite boundaries are restricted to the plane of the film, and other interface is TiN/SiO$_2$ at the bottom of the film. TiN has FCC structure with the atomic distance between (100) planes as 4.2 $\AA$.

The sheet resistance of the film is measured at ambient conditions, and the trasnport properties are shown in panel (a) in Fig. 1. The superconducting transition very close to 1.3 K is seen, which preceeds by a considerable negative slope of resistance vs temperature. The same film was heated at 350 degree, at ambient atmosphere for 10 seconds. The transport measurements in the heated film are shown in panel (b) in Figure 1. It shows the increase in the sheet resistance at 300 K, the increased height of the precursor peak and a subsequent superconducting transition around 1 K. Transport measurements assert that we are studying the identical film only except the level of disorder.  

In Figure 2, the atomic level images in the unheated film (panel (a)) and the heated film (panel (b)) are shown. We observe these images at several places on the film, and it ascertains the conductance behavior measured corresponds to the TiN surface.

\begin{figure}[h]
\includegraphics[width=0.48\textwidth]{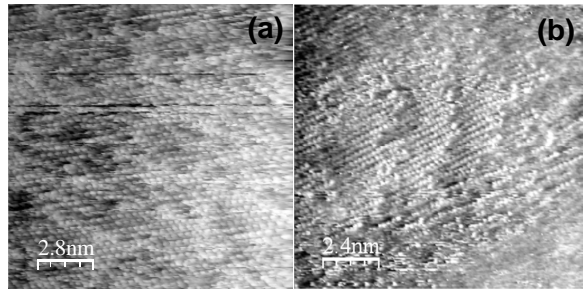}
\caption{Atomic level images in (a) the unheated (a) and (b) the heated films of TiN. The area in the image is around 12 and 10 nm, respectively, and the corresponding lengths are shown at the bottom left corner in the panels.}
\end{figure}

In figure 3, the conductance curves in the unheated and the heated film are shown. The curves are measured at 100 mK, and at zero magnetic field. The superconducting gap in the unheated film is seen in the main panel, which opens up on the reduced conductance close to zero bias voltage. In the heated film, the superconducting gap can be seen in the  bias voltage range from -1 mV to +1 mV as shown in the inset. In bias voltage range from -9 to +9 mV, we only observe the monotonic fall of the conductance as we approach the zero bias. If we calculate the second derivative of the superconducting gap, the estimated size changes from 0.37 mV to 0.143 mV when we heat the same film. This suggests that the heated film has considerably smaller superconducting gap, than the unheated film.    

In figure 4, panels (a) and (b), we show the conductance curves at several magnetic fields, in the unheated and the heated film, respectively. These curves are measured at 100 mK at the same position of the tip. At zero magnetic field, the superconducting gap displays presence of quasi-particle peak positions, and at higher fields these are suppressed. In panel (a), the zero bias conductance evolves upto 4 T, and at 5 T, the conductance nearly remains the same. The rapid change of conductance is from 1.5 T to 2.5 T. In the heated film, the conductance curves at 0 and 0.5 T are identical near zero bias, however, the appearance of quasi-particle peak positions is seen only at 0 T. At 1 T, the zero bias conductance rises, and tend to merge with the conductance behavior shown in the figure 3. In the heated film the positioning of the tip on the sample was difficult at higher magnetic fields, if we choose the smaller bias voltage range for measurement. This limits further evolution of the conductance change, related to the superconducting behavior. 

\begin{figure}[h]
\includegraphics[width=0.48\textwidth]{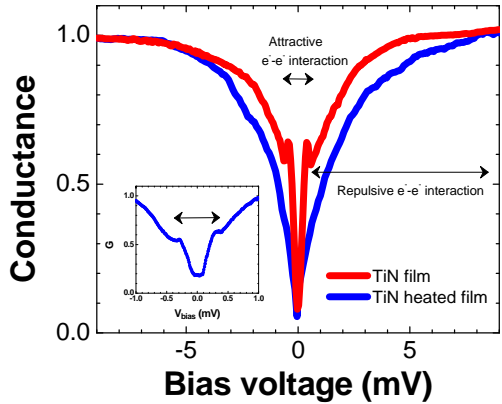}
\caption{\label{fig1} In the main panel we plot the conductance curves at 0 T and 100 mK in (a) the unheated film shown by red curve, and (b) the heated film shown by blue curve. The inset shows a separate measurement in the heated film in the bias voltage range from -1 mV to +1 mV.}
\end{figure}

The unheated film has B$_{c2}$ near 2.5 T, estimated from the transport measurements. The conductance curves at 4 T and 5 T, correspond to the normal state in the film. The suppression of the conductance near zero bias, therefore, is peculiar, and it is related to the Atshuler-Aronov zero bias anamoly due to enhanced electron-electron interaction \cite{Altshuler1,Altshuler2,Bartosch}, in the 2-d disordered films. 

In figure 5, we show, the temperature evolution of the conductance curves, at 0 T and at 7 T, in both the films. The range of bias voltage shown is from -4.5 mV to +4.5 mV. Panels (a) and (c) are at 0 T, however, the unheated film clearly shows the opening of the gap, and we do not observe this feature in panel (c) due to the selected bias voltage range. In panels (b) and (d), the conductances curves at 7 T are shown, and these corresponds to the measurements where we did not observe features of superconducting behavior irrespective of the bias voltage range. In all the panels, the conductance curves for the tempeartures below 1 K are shown in the blue color scheme, while above 1 K are in the pink color scheme. All the curves are normalized at around -4.5 mV in order to facilitate comparisons presented here.

\begin{figure}[h]
\includegraphics[width=0.48\textwidth]{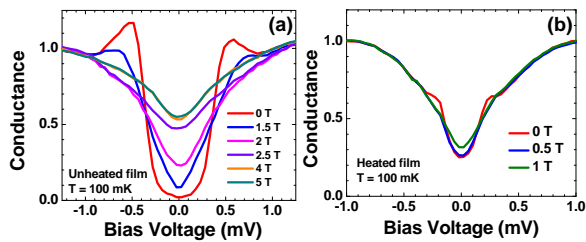}
\caption{\label{fig1} The panel (a) shows the magnetic field dependence of the conductance curves in the unheated film. Selected fields are shown in different colors. In panel (b) the conductance curves at three values of magnetic field are shown. Both sets of measurements are carried out at 100 mK.}
\end{figure}

\begin{figure}[h]
\includegraphics[width=0.48\textwidth]{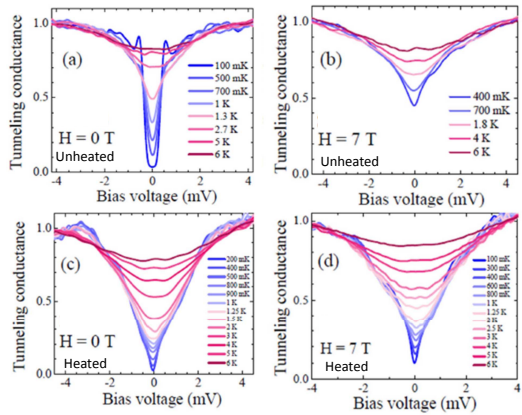}
\caption{\label{fig1} Temperature and magnetic field dependence at 0 T and 7 T, in the unheated film (panels (a) and (b)) and the heated film (panels (c) and (d)). The curves at temperatures below 1 K are in the blue color scheme and above 1 K are in the pink color scheme. The measurement corresponds to the same position of the tip, at stabilized temperatures in the warm up mode.}
\end{figure}

At high fields of 7 T, we only observe the zero bias anamoly \cite{Kulkarni}, starting at around 6 K in both the films. Near and above 6 K, we were unable to resolve the relative changes in the zero bias conductance, and we believe this is uniform for both the films at all magnetic fields. The zero bias conductance at 7 T shows considerable suppression, as the temperature is reduced. As the temperature approaches 1 K, the zero bias anamaloy in the heated film is seen to be stronger than in the unheated film. At 400 mK, the zero bias conductance in the unheated film is around 0.5, while it reaches to 0.2 when the film is heated. This suggests that the film which is heated, and higher sheet resistance, has stronger zero bias anamloy in the normal state, compared to the unheated film. It is therefore reasonable to consider that the superconducting gap opens up on the reduced states in the heated film. In such a case, it is more difficult to form superconducting phase, and the gap size also reduces. This can be asserted from the panel (a), where the two distinct phenomenon seen together at 0 T conductance curves at 100 mK. If one compares the curves in panel (c) and (d), the zero bias anamoly does not change strongly by the magnetic field in the heated film. Our data suggests the normal state in both the film is 2-d disordered metal, and the subsequent superconducting phases are proportionally affected by the presence of the disorder. 

\begin{figure}[h]
\includegraphics[width=0.48\textwidth]{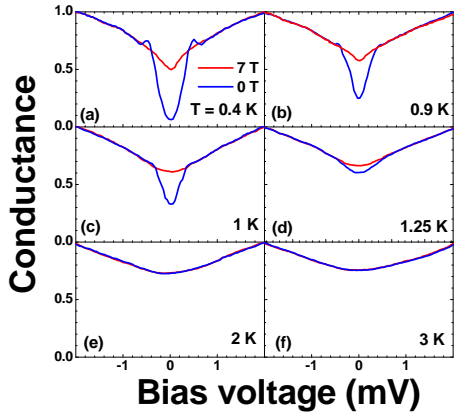}
\caption{\label{fig1} Panels (a) to (f) are the comparative plots of the conductance curves at 0 T and 7 T together, at six different temperatures, four below T$_c$ and two above T$_c$. The measurements correspond to the unheated film, and the respective temperatures are shown in each panel.}
\end{figure}

In order to monitor the closing of the superconducting gap, we plot together, the 0 T and 7 T curves, in the unheated film at few selected temperatures. This comparison is shown in Figure 6, from panels (a) to (f). The conductance curve at 0.4 K, shows the superconducting and the normal state behavior. In the temperature range from 0.9 K to 1.25 K, we observe a clear difference between the zero bias conductance at 0 T and 7 T, shown by blue and the red curve respectively. At 2 K and 3 K, the superconducting gap completely closes, and we do not observe any features of superconducting correlations. We are therefore unable to resolve any finer features related to pseudo-gap in our measurements. 

The comparisons of the conductance curves at different magnetic fields, and temperatures, in the unheated and heated film suggests a possible phase diagram in the disordered TiN films. The phase diagram is shown in Figure 7, where we broadly demark the different regions of the electronic behavior based on our local measurements. The normal state metallic conductivity is observed above 6 K, however the temperature range from 6 K down to the respective superconducting transition temperatures are behaviors related to the 2D-disordered metal. If we start with a film with H,T phase diagram resembling to the one shown with the solid dark line, the increase in the disorder in the same film, represented by the parameter $\delta$, would lead to the shift in the phase boundary in the direction shown by the black arrow.  

\begin{figure}[h]
\includegraphics[width=0.48\textwidth]{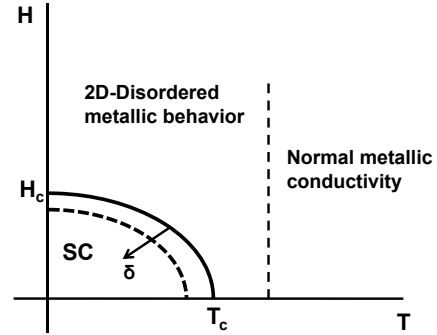}
\caption{\label{fig1} A phase diagram is shown in the figure, demarking three different regions as mentioned. The dashed lines are used as a soft phase boundary. The parameter $\delta$ corresponds to the degree of disorder and is shown in the increasing manner by the direction of the arrow.}
\end{figure}

In conclusion, we associate the normal state characteristics with the Atshuler-Aronov zero bias anomaly due to enhanced electron-electron repulsion in the presence of disorder. We find strong indications that this contribution dominates the local conductance behavior at all temperatures and at high magnetic fields, in contrast to the role of superconducting fluctuations forming a pseudogap behavior as reported earlier. The superconducting state which occurs over an energy scale within ±0.5 mV range, is formed over the disordered 2d-metallic phase with logarithmic conductance behavior. With the heat treatment at ambient the same polycrystalline film shows the increase in the sheet resistance and the decrease of $T_c$ in the transport data, and in STS at 100 mK we found further reduction of density of states close to zero bias at high fields and at zero fields. A superconducting gap was seen at several places, with reduced size, which relates to the depleted normal state density of states at Fermi energy when compared to the unheated film. 

We acknowledge support of UAM workshop SEGAINVEX. This work was supported by the Spanish MINECO (Consolider Ingenio Molecular
Nanoscience CSD2007-00010 program, FIS2011-23488 and Indo-Spain collaboration ACI-2009-0905), by the Comunidad de Madrid through program Nanobiomag-net (S2009/MAT-1726), by the program Quantum mesoscopic and disordered structures of the Russian Academy of Sciences, by the Russian Foundation for Basic Research (Grant No. 12-02-00152 and Grant No. 12-02-31302), and by the US Department of Energy Office of Science under the Contract No. DEAC02- 06CH11357.


\begin{thebibliography}{99}


\bibitem{Strongin70} M. Strongin, R. S. Thompson, O.F. Kammerer, and J. E. Crow, Phys. Rev. B 1, 1078 (1970).

\bibitem{Fisher} M. P. A. Fisher, and D. H. Lee, Phys. Rev. B 39, 2756 (1989).

\bibitem{Haviland} D. B. Haviland et al., Phys. Rev. Lett. 62, 2180 (1989); Y. Liu et al., Phys. Rev. B 47, 5931 (1993).

\bibitem{Hebard} A. F. Hebard, and M. A. Paalanen, Phys. Rev. Lett. 65, 
927 (1990).

\bibitem{Sacepe} B. Sacepe et al., Nature Physics 7, 239 (2011).

\bibitem{Sacepe1} B. Sacepe, C. Chapelier, T. I. Baturina, V. M. Vinokur, M. R. Baklanov, and M. Sanquer, Phys. Rev. Lett. 101, 157006 (2008).

\bibitem{Sacepe2} B. Sacepe, C. Chapelier, T. I. Baturina, V. M. Vinokur, M. R. Baklanov, and M. Sanquer, Nature Communications 1, 140 (2010).

\bibitem{Noat} Y. Noat, T. Cren, C. Brun, F. Debontridder, V. Cherkez, 
K. Ilin, M. Siegel, A. Semenov, H. Hubers, and D. Roditchev ArXiV 1205.3408 (2012).

\bibitem{Baturina} T. I. Baturina et al. Phys. Rev. Lett. 99, 257003 (2007).

\bibitem{Baturina1} T. I. Baturina et al., Pisma v ZhETF 79, 416 (2004) [JEPT Lett. 79, 337 (2004)].

\bibitem{Pfuner} F. Pfuner, L. Degiorgi, T. I. Baturina, V.M. Vinokur, M. R. Baklanov, New J. Phys. 11, 113017 (2009).

\bibitem{Altshuler1} B. Altshuler, A. Aronov, and P. Lee, Phys. Rev. Lett. 44, 1288 (1980).

\bibitem{Altshuler2} B. Altshuler, and A. Aronov, Elsevier, Electron
Electron Interaction in Disordered Conductors (1985).

\bibitem{Bartosch} L. Bartosch, and P. Kopietz, Eur. Phys. J. 28, 29 (2002).

\bibitem{Kulkarni} P. Kulkarni, H. Suderow, S. Vieira, M.R. Baklanov, T. Baturina, V. Vinokur, Arxiv, arXiv:1401.4694.

\end{thebibliography}
\end{document}